\documentclass[twocolumn,floatfix,showpacs,preprintnumbers,amsmath,amssymb]{revtex4}

\usepackage{graphicx}
\usepackage{dcolumn}
\usepackage{bm}

\begin{document}

\title{
Vibrational nonequilibrium effects in the conductance of 
single-molecules with multiple electronic states
}

\author{R. H\"artle}
\author{C. Benesch}
\author{M. Thoss}%
\affiliation{%
Department of Chemistry, Technical University of Munich, Lichtenbergstrasse 4, 
D-85747 Garching, Germany
}%

\date{\today}

\begin{abstract}
Vibrational nonequilibrium effects in charge transport through single-molecule junctions
are investigated. Focusing on molecular bridges with multiple electronic 
states, it is shown that electronic-vibrational coupling triggers a variety of 
vibronic emission and absorption processes, which 
influence  the conductance properties and mechanical stability of single-molecule 
junctions profoundly.
Employing a master equation and a nonequilibrium Green's function approach, 
these processes are analyzed in detail for
a generic model of a molecular junction and 
for benzenedibutanethiolate bound to gold electrodes.
\end{abstract}

\pacs{73.23.-b,85.65.+h,71.38.-k}
\maketitle

\emph{Introduction:}\ Charge transport through single-molecule junctions, i.e.\ molecules which are
bound to metal or semiconductor electrodes, 
represents an interesting and challenging nanoscale nonequilibrium problem.
Recent experimental advances  have allowed to study the conductance properties
of single-molecule junctions and revealed a wealth of intriguing
transport phenomena \cite{Reed97,Reichert02,Xiao04,Parks07}.
An important aspect that distinguishes nanoscale molecular conductors from mesoscopic 
devices  is the  influence of the nuclear degrees of freedom 
of the molecular bridge \cite{Qiu04,Sapmaz06,Galperin07}. 
Due to the small size of molecules, the charging of the molecular bridge is
often accompanied by significant changes of the nuclear geometry that
indicate strong coupling between electronic and nuclear (in particular vibrational) 
degrees of freedom \cite{Pasupathy05}.
Electronic-vibrational (vibronic) coupling 
manifests itself in vibrational structures in the conductance,
which have been observed for a variety of different systems 
\cite{Kushmerick04,Qiu04,Djukic05,Sapmaz06,Thijssen06},
and may result in current-induced vibrational excitation that destabilizes the
junction and causes local heating \cite{Schulze08,Ioffe08}.
Furthermore, conformational changes of the geometry of the conducting molecule are 
possible mechanisms for switching behavior and negative differential resistance
\cite{Choi,Gaudioso00}. 
The physical mechanisms of many of these phenomena, in particular in the resonant transport regime,
are not well understood.
\begin{figure}[b]
\begin{tabular}{ccc}
\begin{tabular}{c}
\resizebox{0.16\textwidth}{0.21\textwidth}{
\includegraphics{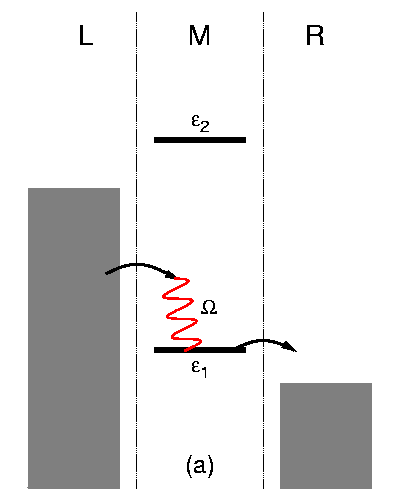}
}
\end{tabular}
&\hspace{-6mm}
\begin{tabular}{c}
\resizebox{0.16\textwidth}{0.21\textwidth}{
\includegraphics{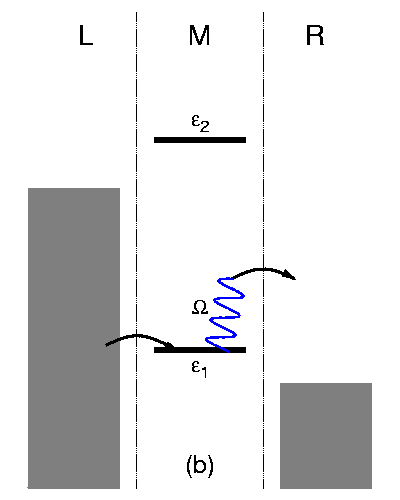}
}
\end{tabular}
&\hspace{-6mm}
\begin{tabular}{c}
\resizebox{0.16\textwidth}{0.21\textwidth}{
\includegraphics{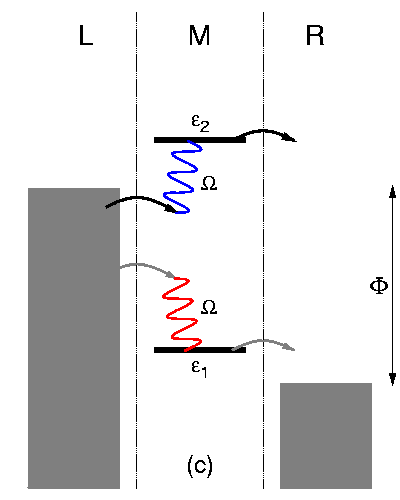}
}
\end{tabular}
\\
\end{tabular}
\caption{\label{figEmiAb} Schematic representation of vibronic transport processes in a molecular junction
involving
emission (a) and absorption (b) of vibrational quanta upon transmission of an electron through  
a single electronic state, as well as emission and subsequent absorption associated  with sequential electron transmission via two different electronic states (c). 
}
\end{figure}

In this paper, we investigate vibrational effects in 
resonant electron transport through single-molecule junctions
with multiple electronic states. The basic mechanisms of this nonequilibrium problem
are illustrated in Fig.\ \ref{figEmiAb}. 
The interaction of a transmitted electron  with the vibrational degrees of freedom of
the central molecule may result in excitation of vibrational quanta (in the following referred to
as emission process, Fig.\ \ref{figEmiAb} (a)), or deexcitation (in the following referred to
as absorption process, Fig.\ \ref{figEmiAb} (b)). For molecular junctions
that are dominated by transport through a single electronic state, these basic processes and the resulting conductance properties
have been studied in detail employing a variety of theoretical methods including
 master equations, scattering theory, nonequilibrium Green's
function approaches and path integral methods 
\cite{Galperin07,May02,Lehmann04,Cizek04,Mitra04,Wegewijs05,Ryndyk06,Semmelhack,Muehlbacher08}. 

The situation is considerably more complex if
multiple electronic states of the molecular bridge are involved in the transport process. In particular,
higher lying electronic states can facilitate deexcitation of the 
vibrations via resonant absorption processes associated with the sequential transmission of two electrons (Fig.\ \ref{figEmiAb} (c)), thus  stabilizing 
the molecular junction.
Moreover, vibronic coupling results in an effective interaction
of  electrons in different electronic states similar to electron-electron coupling 
in Hubbard-like models. 
As a result of the intricate interplay of these different processes 
vibrational and electronic  signatures in the conductance of a single molecule  may become 
of the same importance.

\emph{Theory:}\ To study vibrationally coupled electron
transport in molecular junctions, we consider  a single molecule that
is covalently bound to two metal leads (L,R)  described by the Hamiltonian
$H$=$H_{\text{el}}$+$H_{\text{vib}}$. The electronic part of the Hamiltonian, 
\begin{equation}
H_{\text{el}} = \sum_{i\in\text{M}} \epsilon_{i} c_{i}^{\dagger}c_{i} + 
\sum_{k\in\text{L,R}} \epsilon_{k} c_{k}^{\dagger}c_{k} + \sum_{k,i} ( V_{ki} 
c^{\dagger}_{k} c_{i} + \text{h.c.} ),
\end{equation}
involves the electronic states of the molecular bridge with 
energies $\epsilon_{i}$ that are  coupled by interaction matrix elements $V_{ki}$ 
to electronic states in the leads with  energies $\epsilon_{k}$.
The vibrational degrees of freedom of the molecular bridge are described by
\begin{equation}
\label{Hvib}
H_{\text{vib}} = \sum_{\alpha} \Omega_{\alpha} 
a_{\alpha}^{\dagger}a_{\alpha} + \sum_{\alpha,i\in\text{M}} \lambda_{i\alpha} 
(a_{\alpha}+a_{\alpha}^{\dagger}) c_{i}^{\dagger}c_{i}. 
\end{equation}
Here, $a_{\alpha}$ is the annihilation operator for a vibration with frequency $\Omega_{\alpha}$, 
and $\lambda_{i\alpha}$ denote the corresponding vibronic coupling constants.

Several observables are of interest to
analyze the interplay of the electronic and vibrational degrees of freedom in nonequilibrium 
transport through a molecular junction. Here, we consider specifically
the current-voltage characteristics $I(\Phi)$ and the vibrational 
excitation $\langle a_{\alpha}^{\dagger} a_{\alpha}\rangle$.
To calculate these observables, we employ two complementary approaches: 
a nonequilibrium Green's function (NEGF) method 
\cite{Hartle,Galperin06} as well as a master equation (ME) approach. 
Briefly, both approaches are based on 
the small polaron transformation of the Hamiltonian \cite{Galperin07},
which facilitates a 
nonperturbative description of vibronic interaction. The transformed Hamiltonian $\overline{H}$ 
comprises an exactly solvable part $\overline{H}_{0}$ and a renormalized molecule-lead 
coupling term $\overline{V}=\sum_{k,i} ( V_{ki} X_{i} 
c^{\dagger}_{k} c_{i} + \text{h.c.} )$, which involves the shift operators $X_{i}=\text{exp}(\sum_{\alpha}(\lambda_{i\alpha}/\Omega_{\alpha})(a_{\alpha}-a_{\alpha}^{\dagger}))$. 
It is noted that for models with multiple electronic
states  $\overline{H}_{0}$ includes
Hubbard-like terms $\sim \lambda_{i\alpha}\lambda_{j\alpha}c_{i}^{\dagger}c_{i}c_{j}^{\dagger}c_{j}$, 
$i$$\neq$$j$, which describe vibrationally mediated electron-electron coupling. 

In the ME approach, all observables are obtained from the reduced density 
matrix $\rho$ of the electronic and vibrational degrees of freedom of the molecule,  which  
is given as the  stationary limit of the well-established 
equation of motion \cite{Mitra04,Lehmann04}
\begin{eqnarray}
\label{redmatEOM}
0 =  - i \left[ \overline{H}_{0} , \rho \right] - \int_{0}^{\infty}\text{d}\tau\,
\text{tr}_{\text{leads}}\bigl\lbrace\bigl[\,\overline{V},\left[\,\overline{V}(\tau),\rho\, \rho_{\text{leads}}^{\text{eq}}\right]\bigr]\rbrace.
\end{eqnarray}
Here, $\rho_{\text{leads}}^{\text{eq}}$ denotes the equilibrium density matrix of the leads. In the results
reported below, electronic coherences of the density 
matrix $\rho$ are fully taken into account, while
vibrational coherences are negligible. 
The current
is given by 
$I=(2e/\hbar)\int_{0}^{\infty}\text{d}\tau\hspace{1mm} \text{tr}\lbrace \left[\overline{V}(\tau),\rho \rho_{\text{leads}}^{\text{eq}}\right] \hat{I}\rbrace$,
with 
$ \hat{I} = \sum_{i,k\in\text{L}}  ( V_{ki} c_{k}^{\dagger}c_{i}X_{i} - \text{h.c.} )$ 
and $\overline{V}(\tau)=\text{e}^{-i\overline{H}_{0}\tau} V \text{e}^{i\overline{H}_{0}\tau}$.

The  NEGF method is an extension of the approach outlined in Refs.\ \onlinecite{Hartle,Galperin06}
to treat multiple electronic states. Thereby, the electronic Green's 
function matrix $\textbf{G}$ is determined by the self energy matrix 
$\mathbf{\Sigma}_{\text{L/R},ij}(\tau,\tau')=\sum_{k\in\text{L/R}}V_{ki}^{*}V_{kj}g_{k}(\tau,\tau')\langle 
\text{T}_{c} X_{j}(\tau')X^{\dagger}_{i}(\tau)\rangle$, where $g_{k}$ 
denotes the free Green's function 
of lead state $k$ and $T_{c}$ denotes time-ordering on the Keldysh contour. 
Vibrationally induced electron-electron interaction is treated non-perturbatively 
following the strategy of Ref.\ \onlinecite{Groshev}. 
The calculation of the correlation functions $\langle 
\text{T}_{c} X_{j}(\tau')X^{\dagger}_{i}(\tau)\rangle$ is based on a cumulant expansion 
in the dimensionless coupling parameters $\lambda_{i\alpha}/\Omega_{\alpha}$, which in turn
involves  the electronic Green's function and, therefore, is obtained
in a self-consistent scheme \cite{Galperin06,Hartle}. Within the NEGF method, the current is  given by
$I=(2e/\hbar)\int\left(\text{d}E/(2\pi)\right)\hspace{1mm} \text{tr}\lbrace \mathbf{\Sigma_{\text{L}}^{<}}\textbf{G}^{>}-\mathbf{\Sigma_{\text{L}}^{>}}\textbf{G}^{<} \rbrace$.

In both approaches the molecule-lead coupling is treated within (self-consistent) 
second order perturbation theory. 
The ME approach includes all resonant processes, however, it neglects 
contributions related to 
co-tunneling, and does not account for the broadening of the 
electronic levels induced by the interactions with the leads. These processes
are taken into account in the NEGF method \cite{Galperin06,Hartle}, which, as a result, becomes exact for
vanishing vibronic coupling. 
On the other hand, the ME approach allows an exact treatment  of the Hubbard-like terms 
in $\overline{H}_{0}$, while these are approximately accounted for in the NEGF method.
Thus, both approaches are complementary and the good agreement of the results found below
is a strong indication of their validity.


\emph{Results:}\ First, we consider a generic model 
system for a molecular junction, represented by two electronic levels 
and a  single vibrational mode with frequency $\Omega$=0.1 eV.
Both leads are modeled by semi-elliptic 
bands \cite{Cizek04} with a band width of 3 eV and an overall molecule-lead
coupling strength of  0.1 eV. 
Throughout this article, we consider a low temperature of  $T=10$ K to 
ensure that vibrational excitation is caused solely by nonequilibrium processes.

\begin{figure}
\resizebox{0.4\textwidth}{0.26\textwidth}{
\includegraphics{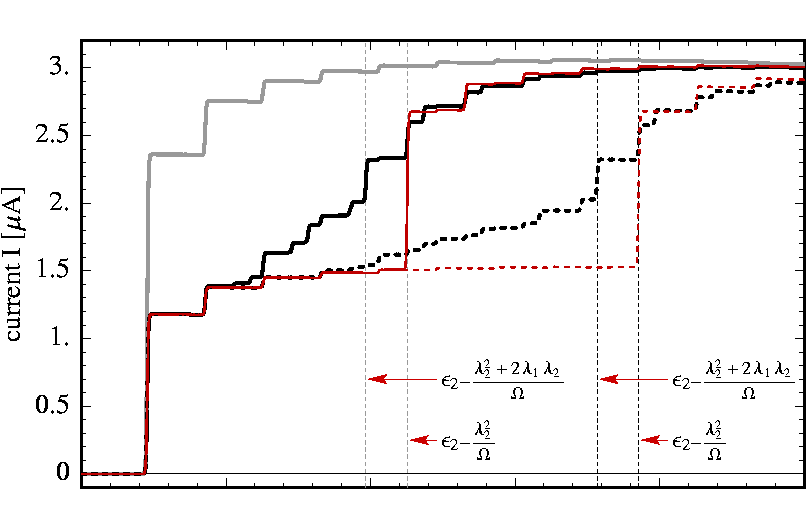}
}
\vspace{-0.35cm}\\ \hspace{-1.7mm}
\resizebox{0.402\textwidth}{0.26\textwidth}{
\includegraphics{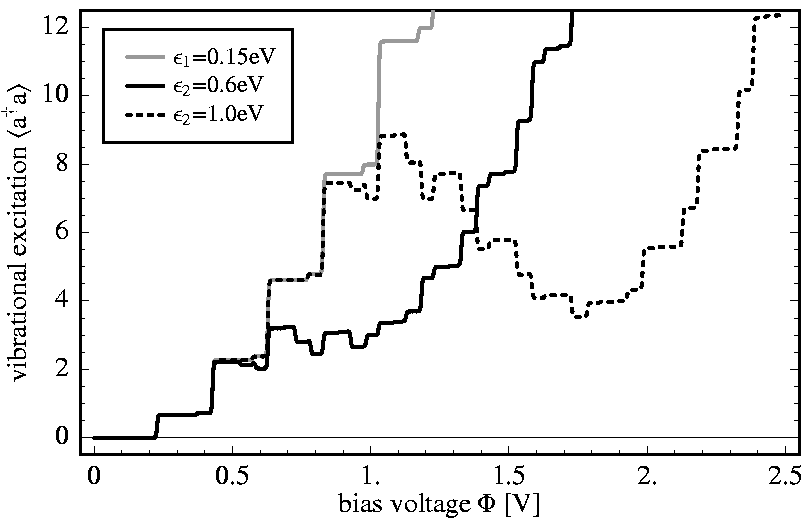}
}
\caption{\label{figME42} Current and vibrational excitation for a generic model 
system with two electronic levels and a single vibrational mode obtained for two different locations of
the second electronic state. The red lines depict results of an approximate treatment where the
current through the two electronic states has been calculated separately and the grey lines give  results
for a single electronic state.}
\end{figure}

Fig.\ \ref{figME42} shows the current-voltage characteristic and the average vibrational
excitation for a moderate vibronic coupling, 
$\lambda_{1/2}$=0.06 eV, obtained with the ME approach.  
The first electronic level is  located at 
$\epsilon_{1}$=0.15 eV (relative to the Fermi energy
of the leads)
and we consider two different locations
of the second electronic level, $\epsilon_{2}$.

As a reference, also the result for a single electronic level at $\epsilon_{1}$ 
is shown.
For the latter result, the molecule-lead  coupling has been enhanced by a factor $\sqrt{2}$ to obtain
the same current as in the two-level cases for large voltages. Overall, the interaction of the transmitted electrons with the vibration results in
pronounced vibronic resonance structures in the current-voltage characteristic and 
significant vibrational excitation.
As has been previously studied in detail for a single electronic level \cite{Hartle,Galperin06}, the current
exhibits a steplike increase  at 
voltages $\Phi\approx 2(\epsilon_{1}-\lambda_{1}^{2}/\Omega+n\Omega)$, 
$n=0,1,2,..$. 
The corresponding current-induced vibrational excitation also shows
a step-like monotonous increase with voltage, and acquires considerable values.

The situation becomes significantly more complex if another electronic level participates in the transport process. 
In the current voltage-characteristic, this results in a wealth of additional resonance structures. 
Due to vibrationally  induced electron-electron 
interaction, the resonance, which for a purely electronic transport mechanism
would appear at  $\Phi \approx 2\epsilon_{2}$, splits into two resonances at
$\Phi \approx 2(\epsilon_{2}-\lambda_{2}^{2}/\Omega)$ and 
$\Phi \approx 2(\epsilon_{2}-(\lambda_{2}^{2}+2\lambda_{1}\lambda_{2})^{2}/\Omega)$ (indicated by the 
dashed lines in Fig.\ \ref{figME42}). 
They correspond to the first electronically excited state of the anion and the 
electronic ground state of the di-anion, respectively \cite{splittings}.
Vibrational progressions originating from all three electronic resonances towards higher voltages 
can be seen. An intriguing effect is the appearance of 
pronounced vibronic resonance structures at lower voltages, where the
electronic level $\epsilon_{2}$ is still outside the conductance window.
These structures are related to the absorption processes schematically depicted in Fig.\ \ref{figEmiAb}(c): 
Upon transmission via the lower lying electronic resonance $\epsilon_{1}$, an electron may 
excite vibrational quanta. A subsequently  transmitted electron can absorb these quanta, thus effectively lowering
the energy of the resonances. This results in vibronic structures at voltages  
$\Phi\approx 2(\epsilon_{2}-\lambda_{2}^{2}/\Omega-n\Omega)$ and 
$\Phi\approx 2(\epsilon_{2}-(\lambda_{2}^{2}+2\lambda_{1}\lambda_{2})^{2}/\Omega-n\Omega)$, 
$n=0,1,2,..$. As a consequence of the additional vibronic structures, the purely electronic
resonance steps are barely visible.
We emphasize that this resonant absorption process is a collective effect of the two 
electronic states and
cannot be described in an approximate treatment, which calculates the currents 
through the two electronic levels separately (red lines in Fig.\ \ref{figME42}). 

The absorption of current-induced vibrational energy via higher-lying electronic states
has an even more pronounced effect on the
vibrational nonequilibrium distribution in the stationary state. The results in
Fig.\ \ref{figME42} demonstrate that  this process
may reduce the vibrational energy by more than 50\%. 
Thus, higher lying electronic levels stabilize single-molecule
junctions with respect to local heating. 

\begin{figure}
\resizebox{0.4\textwidth}{0.26\textwidth}{
\includegraphics{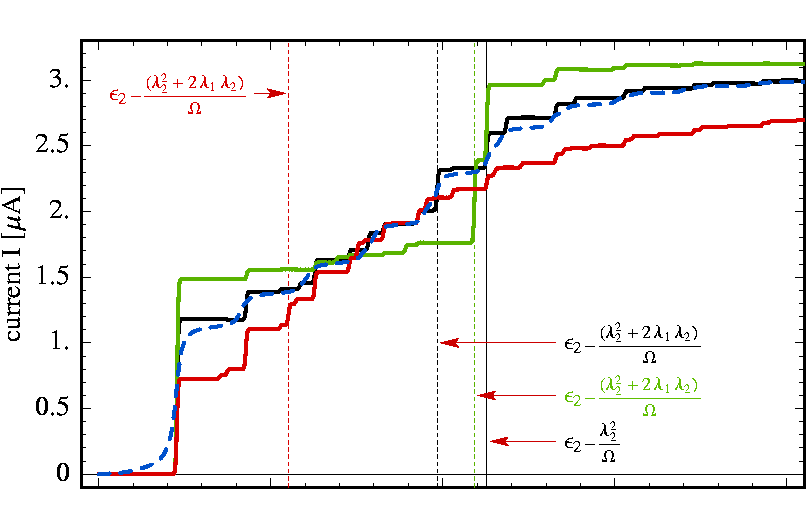}
}
\vspace{-0.35cm}\\ \hspace{-2.1mm}
\resizebox{0.4\textwidth}{0.26\textwidth}{
\includegraphics{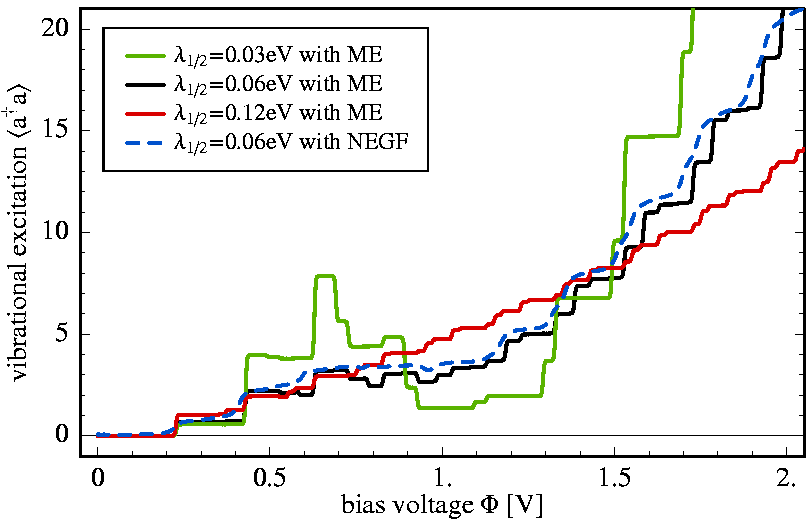}
}
\caption{\label{figdegphon} Current and vibrational excitation for a generic 
model system with two electronic states, $\epsilon_{1}$=0.15 eV and $\epsilon_{2}$=0.6 eV,
and different vibronic coupling parameters as indicated in the legend.
}
\end{figure}

Another interesting aspect is the influence of the vibronic coupling strength  $\lambda_{1/2}$.
Fig.\ \ref{figdegphon} shows results obtained 
for the two-state model considered above,
but different vibronic coupling parameters. 
Thereby,  the energies of the electronic states $\epsilon_{1/2}$ have been adjusted to 
compensate for the different polaron shifts, $\lambda_{1/2}^{2}/\Omega$. 
In the current-voltage characteristic,
a larger vibronic coupling results in more pronounced vibrational step structures.
The results obtained for the vibrational excitation
show that, except for regions with  resonant absorption, the largest current-induced vibrational 
excitation is obtained for the system with the weakest vibronic coupling.
This counterintuitive behavior has been analyzed before for the case
of a single electronic state \cite{Semmelhack,Ryndyk06}.
In particular, it has been shown that the vibrational excitation diverges in the limit 
$\lambda$\hspace{0mm}$\rightarrow$0 if no other vibrational relaxation processes are active. 
The present results demonstrate that this vibrational instability can be considered as an 'artifact' 
of the restriction to a single electronic level on the molecule. 
Vibrational absorption processes related to higher-lying electronic states, 
which will always be present in a real molecule, limit the vibrational excitation,
and thus, prevent this  instability. 

The comparison of results obtained with the ME approach 
and NEGF theory in Fig.\ \ref{figdegphon} reveals very good agreement for small (data not shown) 
and moderate  vibronic coupling, and thus, indicates the validity
of the two approaches. The only major difference is the  broadening of 
the step structures due to tunneling
processes. For strong vibronic coupling, e.g.\ $\lambda_{1/2}=0.12$ eV, 
the NEGF approach ceases to be valid due to the cumulant expansion employed \cite{Hartle,Galperin06}.  

\begin{figure}[t]
\resizebox{0.4\textwidth}{0.26\textwidth}{
\includegraphics{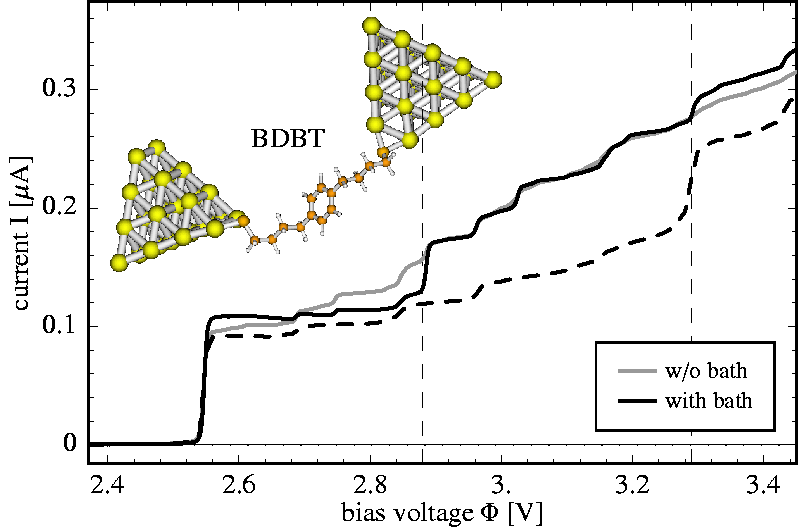}
}
\caption{\label{figBDBT} Current-voltage characteristic 
for a BDBT molecular junction, with and without coupling to a  bath. 
The dashed line depicts results of an approximate treatment \cite{Hartle}, where the
current through the two electronic states has been calculated separately.}
\end{figure}

The results discussed so far were obtained for a generic model system of a molecular junction. 
Finally, we consider vibrational nonequilibrium effects in
electron transport through a realistic molecular junction, 
where a p-benzene-di(butanethiolate) molecule (BDBT) is 
covalently bound to  two gold electrodes. Recent first-principle electronic
structure calculations for this system have shown that the transport properties of
this junction can be well represented
by a model that includes two electronic states, localized on the molecular bridge,
 and four vibrational modes \cite{Benesch08b}.
The influence of the remaining vibrational degrees of freedom of the
molecule and the phonons of the gold electrodes are represented by a coupling to a bath 
as described in Ref.\ \onlinecite{Hartle}.
It is noted that, in contrast to the generic model
discussed above, the two molecular levels that dominate charge transport through BDBT 
are occupied in equilibrium, and thus, the prevailing mechanism is hole transport. Correspondingly, 
the electronic-vibrational coupling term in Eq.\ \ref{Hvib} is replaced by $\sum_{\alpha,i\in\text{M}} (-\lambda_{i\alpha}) 
(a_{\alpha}+a_{\alpha}^{\dagger}) c_{i}c_{i}^{\dagger}$.

Fig.\ \ref{figBDBT} shows the respective current-voltage characteristics 
using NEGF theory. 
Similar as for the model system discussed above, the presence of two vibrationally-coupled 
electronic states results in a splitting of the resonance corresponding to the lower electronic state
into two step structures (indicated by the dashed vertical lines). The vibronic coupling
causes significant current-induced  vibrational excitation
and manifests itself in the current-voltage characteristic in a wealth of 
vibrational structures due to emission and absorption processes.
As a consequence, electronic and vibrational features are hardly distinguishable.
This scenario might be very common for realistic molecular junctions, which typically involve many 
active vibrational modes. 
If the coupling to the bath is included (solid black line), 
vibrational relaxation processes compete with current-induced excitation resulting in an 
overall smaller vibrational excitation.  As a consequence, 
resonant absorption processes are less important and electronic resonances in the current-voltage 
characteristic are more pronounced.

\emph{Conclusions:}\ 
Processes due to electronic-vibrational coupling can influence 
the conductance properties and mechanical stability of single-molecule junctions profoundly.
The results presented in this paper demonstrate that in molecular junctions where multiple 
electronic states  participate in the nonequilibrium transport, a number of additional vibronic
processes have to be considered.
These include, in particular, resonant absorption processes associated with the sequential transmission of two electrons
as well as  vibrationally induced effective electron-electron  interaction. 
While the latter  results in a splitting of electronic resonances, the former 
may cause a wealth of additional vibronic structures in the current-voltage characteristic. Moreover,
the resonant absorption process  facilitates 
vibrational cooling and thus represents an important stabilization mechanism of molecular junctions.
Since polyatomic molecules include numerous vibrational modes and often exhibit multiple
closely-lying electronic states, these findings are expected to be of relevance for
most molecular junctions.

\emph{Acknowledgement:}\ We thank Martin Cizek, Maxim Gelin and Wolfgang Domcke 
for helpful discussions, the Deutsche 
Forschungsgemeinschaft and the German-Israeli Foundation for Scientific 
Development (GIF) for support and  the Leibniz 
Rechenzentrum, Munich, for generous allocation of computing time.



\begin{thebibliography}{28}
\bibitem[{spl()}]{splittings}
\bibinfo{note}{We note that the resonance at $\Phi=2\epsilon_{1}$ may also
  split into two resonances if
  $\vert\epsilon_{1}-(\lambda_{1}^{2}+2\lambda_{1}\lambda_{2})/\Omega-\epsilon%
_{\text{F}}\vert >
  \vert\epsilon_{2}-\lambda_{2}^{2}/\Omega-\epsilon_{\text{F}}\vert$.}
\end{thebibliography}
\end{document}